# Dark Matter in Universe as the Geometry of Empty Space

A. Z. DOLGINOV

Rensselaer Polytechnic Inst., Troy, NY, 12180, USA
adolgi1@yahoo.com

ABSTRACT

The observed excess of gravitational forces in galaxies and galactic clusters is usually referred as the existence of "dark matter particles" of unknown origin. An alternative explanation of the dark matter effect is presented here by assuming that the empty space of the Universe has a complicated geometrical structure. It is assumed that the cosmological $\Lambda$ term, which is approximately constant at scales of several Mpc and describes the dark energy, is not constant at scales of several tenth of Kpc and describes the "dark matter effect". This term is mathematically analogous, but not identical, with the energy-momentum tensor of particles ensembles. In this connection the general expression of the Riemann tensor, which depends on the matter distribution in space and on the empty space geometry, is considered. The Ricci tensor does not describe all possible empty space structures. The tensor of the spin curvature also has to be taken into account. Local deviations of the empty space geometry of Universe from a flat one provide the same observational effect as clouds of invisible particles which demonstrate themselves by their gravitational fields only. The matter, which forms stars and galaxies, will concentrate in gravitational depressions of empty space. The gravitation provides correlation of the luminous matter distribution and geometrical structures of space. The presented theory is compared with observations.

Keywords: Dark matter theory; gravitation; cosmology.

## 1. The geometry of empty space

Astronomical observations show that gravitational fields created by luminous matter in galaxies are not sufficiently large to explain the motions of stars at galaxies peripheries. Also observations of deflection of light from remote stars, when the light crosses some galaxy on its way to the Earth (lensing effect), show stronger gravitational field of this galaxy than what would be created by luminous matter alone [1]. These effects are usually explained as a presence of "dark matter particles" of unknown origin. Such interpretation of observations meets many difficulties because appropriate particles were not found up to now [1].

The reasonable question one can ask - if the gravitation is the only property of the dark matter, why it is assumed that the dark matter consists from some kind of particles? It is an equally reasonable assumption that the dark matter gravitation is not connected with any sort of material substances, but rather a result of the complicated geometry of the empty space.

Gravitational fields of particles ensemble are determined by the energy momentum tensors $T_{ik}$ of the ensemble. However, particles interact with each other not only by gravitation, but also by other forces and the knowledge of $T_{ik}$ is not enough to describe the ensemble.

The gravitational relief of space, in other words space geometry, is completely described by the Riemann tensor. Mathematical structures, similar to the energy-momentum tensors, but not related to any sort of particles can determine the part of the Riemann tensor which determine the empty space geometry.

It will be shown that the additional term in Einstein equation which is needed to describe the "dark matter effect" can be completely interpreted as describing local deflections of the empty space geometry from a flat one.

Apparently this possibility did not attract due attention because the condition that the Ricci tensor is zero, $R_{km} = 0$, disregarding the structure of the Riemann tensor $R_{iklm}$, was considered by many authors as a complete description of empty space geometry. We will start with the general description of such geometry.

a) The geometry of the empty space can be different from the flat one [2]. Non flat metrics of Friedman, De-Sitter, etc are particular cases of the global non flat geometry. Gravitational waves are examples of small scale structures of empty space. There are no theoretical or observational objections on the existence of long lived middle and small scale gravitational structures of the empty space.

b) The Riemann tensor $R_{iklm}$ presents the most general information on the geometry of space and, hence, on the gravitational forces acting in space. The Ricci tensor $R_{km} = g^{il}R_{iklm}$, as well as the scalar $R = g^{km}R_{km}$, contains less information. The Riemann tensor is zero only for flat geometry. The Riemann tensor has fourteen independent components; the Ricci tensor has ten independent components. The $R_{km} = 0$ does not mean that $R_{iklm} = 0$. The most general expression of the Riemann tensor which is: (a) linear with respect to energy momentum tensors $T_{km}$, (b) for which the corresponding Ricci tensor determines Einstein equation and (c) which is not zero for the non-flat geometry of the empty space, has the form:

$$R_{iklm} = \frac{1}{2}\{\frac{\partial^2 g_{im}}{\partial x^k \partial x^l} + \frac{\partial^2 g_{kl}}{\partial x^i \partial x^m} - \frac{\partial^2 g_{il}}{\partial x^k \partial x^m} - \frac{\partial^2 g_{km}}{\partial x^i \partial x^l}\} + g_{np}\{\Gamma^n_{kl}\Gamma^p_{im} - \Gamma^n_{km}\Gamma^p_{il}\} = \frac{4\pi G}{c^4}\{g_{km}T_{il} - g_{im}T_{lk} + g_{il}T_{km} - g_{kl}T_{im} - \frac{2}{3}(g_{il}g_{km} - g_{kl}g_{im})T\} + (R_{iklm})_0 \quad (1)$$

Here $(R_{iklm})_0$ is that part of the Riemann tensor which describes the geometry of the empty space.

$$(R_{iklm})_0 = -\frac{1}{3}(g_{il}g_{km} - g_{kl}g_{im})\Lambda + ... \quad (2)$$

Here the "three dots" denote additional terms that may be present. The $\Lambda$ is, so called, cosmological constant. The $\Lambda$ term, which is usually estimated on scales larger than several Mpc, has to be considered as a part of $(R_{iklm})_0$.

Using the expression (1) we obtain the Ricci second rank tensor as:

$$R_{km} = g^{il}R_{iklm} = \frac{8\pi G}{c^4}\{T_{km} - \frac{1}{2}g_{km}T\} + (R_{km})_0 \quad (3)$$

$$(R_{km})_0 = (g^{il}R_{iklm})_0 = -g_{km}\Lambda + ...$$

The Lambda term in (2) and (3) is usually interpreted [3] as describing the vacuum energy density $\varepsilon_\Lambda = (8\pi G)^{-1}c^4\Lambda$, the vacuum mass density $\rho_\Lambda = \varepsilon_\Lambda c^{-2}$, and the pressure $P = -\varepsilon_\Lambda$ in Universe. Observations show that the large scale geometry of Universe is approximately flat. Estimated $\Lambda$ value is $|\Lambda| \leq 10^{-55} cm^{-2}$, i.e. $\varepsilon_\Lambda = 10^{-7} erg/cm^3$ and $\rho_\Lambda = 10^{-29} g/cm^3$. The average density of ordinary matter in Universe is equal to $\rho = 10^{-30} g/cm^3$.

The empty space is usually considered as a quantum vacuum with virtual particles, that appear and disappear, or a space with some exotic substances as "quintessence" etc. We do not consider this problem here.

Observations of the red shift in the spectra of remote galaxies lead to the conclusion that at large scales, of several hundred Mpc, the Lambda term is approximately constant. However, there are no observational evidences that the vacuum energy density and, hence, the $\Lambda$ values are constant at scales of several hundred Kpc. In particular, it is possible that expressions for $(R_{iklm})_0$ and $(R_{km})_0$, related to small scales of the empty space, are characterized also by tensors $\Lambda_{ik}$, analogous to energy momentum tensors $T_{ik}$ of ordinary matter. The $(R_{iklm})_0$ can include also some other more complicated functions. Gravitational fields have to be described by $T_{ik}$ if the fields are created by some sort of particles or by $\Lambda_{ik}$, including $\Lambda$, if this field is a manifestation of the empty space structures.

Lambda terms, as functions of space and time, can describe the dark matter effect (for small space scales) and the dark energy (for large scale). It is possible that in local regions the Lambda terms could have a value, and even a sign, different from that averaged value over large regions. We can write:

$$(R_{iklm})_0 = g_{km}\Lambda_{il} - g_{im}\Lambda_{lk} + g_{il}\Lambda_{km} - g_{kl}\Lambda_{im} - \frac{2}{3}(g_{il}g_{km} - g_{kl}g_{im})\Lambda + ...$$

(4)

$$(R_{km})_0 = -g_{km}\Lambda + 2\Lambda_{km} + ...$$

The general form of the term $(R_{km})_0$ was usually not included in the Einstein equation, but there are no reasons for assumptions that $(R_{km})_0$ is zero or contains only the constant $\Lambda$ term if $(R_{iklm})_0 \neq 0$.

The non-flat local metric can completely mimic a cloud of hypothetical dark matter which interacts with the luminous matter by gravitational forces only. For example, the corresponding $\Lambda$ term in (2) and (3), which mimics a homogeneous dark matter cloud with the mass density $\mu$, is approximately equal to $\Lambda = -4\pi G\mu c^{-2}$. Here $\mu$ is not considered as the mass density of the cloud, but as a function which characterized the space geometry and, hence, the gravitational field there. Outside such regions the $\Lambda$ will tend to approximately constant value.

The matter in Universe has apparently a fractal structure at least at the scale less than $3.10^8$ light years (stars, galaxies, galactic clusters etc.). There are no factors which prevents existing similar structures of the small and middle scale empty space geometry. However the origin and creation of the space gravitational relief is out of the scope of this article.

## 2. The spin-curvature of space

In the previous section the simplest assumptions on the $(R_{iklm})_0$ explicit expressions were used. Some general properties of equations, which determine the space geometry, are discussed below.
Only a few precise solutions of the nonlinear Einstein equation were obtained up to now. Solutions of equation (1) were also not presented. However, it is important to emphasize that equation (3) is less general than equation (1) and contains less information about the space geometry. In particular, using equation (3), we can lose some information which is related to anti-symmetric properties of the Riemann tensor and, hence, some information on possible gravitational structures.
The $R_{km}$ is a symmetric tensor. It is possible, by converting $R_{iklm}$, construct an anti-symmetric second rank tensor $K_{lm}$, taking into account that not only scalars, vectors and tensors, but also spinors and spin-tensors are classical geometrical functions presenting the group of rotation. The covariant derivative $\nabla_n$ of a spinor $\psi$ is [4]:

$$\nabla_n \psi = \psi_{,n} + \Gamma_n \psi, \qquad \Gamma_n = (1/4)\gamma^m \gamma_{m;n} = \frac{1}{4}\{h^m_{(\alpha)} \frac{dh_{m(\beta)}}{dx_n} - h^m_{(\alpha)} h_{l(\beta)} \Gamma^l_{mn}\}\sigma^{(\alpha\beta)}$$

$$\nabla_m \gamma^n = \gamma^n_{;m} + \Gamma_m \gamma^n - \gamma^n \Gamma_m = 0 \qquad (5)$$

$$g^{ik} = (1/2)(\gamma^i \gamma^k + \gamma^k \gamma^i), \qquad \sigma^{ik} = (1/2)(\gamma^i \gamma^k - \gamma^k \gamma^i)$$

$$0 = \nabla_\beta(\nabla_\alpha \gamma_\mu) - \nabla_\alpha(\nabla_\beta \gamma_\mu) = R_{\mu\lambda\alpha\beta}\gamma^\lambda + [K_{\alpha\beta}, \gamma_\mu]$$

Here $\gamma^k(x) = h^k_\alpha(x)\gamma^\alpha$, where $\gamma^k(x)$ - are Clifford matrices which depend on space-time coordinates. Matrices $\gamma^\alpha$ are Clifford matrices for a flat space-time and $h^k_\alpha(x)$ are tetrads which determine the metric tensor $g^{km}(x)$ and the spin tensor $\sigma^{ik}(x)$.

$$g^{km}(x) = h^k_\alpha(x) h^m_\alpha(x) \qquad \sigma^{km}(x) = h^k_\alpha(x) h^m_\beta(x) \sigma^{\alpha\beta} \qquad (6)$$

Greek indices determine $\gamma^\alpha$, $\sigma^{\alpha\beta}$ etc for the flat geometry. Matrices $\gamma^k(x)$, $\sigma^{ik}(x)$ etc. are used in quantum theory but, by themselves, they are classical quantities which are determined by the space geometry.

Using expression (5) for the gamma matrix, one obtains the matrix tensor $K_{ik}$ as [5]:

$$K_{ik} = \frac{1}{4}\sigma^{lm} R_{iklm} = \frac{\partial \Gamma_i}{\partial x^k} - \frac{\partial \Gamma_k}{\partial x^i} + \Gamma_k \Gamma_i - \Gamma_i \Gamma_k \qquad (7)$$

The spin curvature was usually considered in literature [5] in connection with the electron spin properties but not as a tensor complimentary to the tensor Ricci.

Using expressions (1) and (6) we can present the spin curvature tensor as:

$$K_{lm} = \frac{2\pi G}{c^4}\{\sigma_{nm} T^n{}_l - \sigma_{nl} T^n{}_m - \frac{2}{3}\sigma_{lm} T\} + \frac{1}{4}(\sigma^{ik} R_{iklm})_0 \qquad (8)$$

Tensors $R_{il}$ and $K_{lm}$ are expressed by different combinations of $R_{iklm}$ components, are dependent on the different set of energy-momentum tensors, determine different set of tetrads $h^k{}_\alpha$ and, hence, describe different cases of the space geometry. The $K_{lm}$ includes information on the space structure supplementary to $R_{km}$. The $R_{il} = 0$ does not mean that $K_{lm}$ and $R_{iklm}$ are zero. Taking into account Bianchi's relations and equations (5) one obtains equation for $K_{lm}$. It is important to note that this equation have the form similar to Maxwell equations [5]:

$$\nabla_l K_{ik} + \nabla_i K_{kl} + \nabla_k K_{li} = 0, \qquad \nabla_n K^{nm} = J^m \qquad (9)$$

where $J^m$ determines the divergence-free vector, similar to the electric current.

$$J^m = \frac{4\pi G}{c^4}\sigma^{ik} T^m{}_{i;k} + (J^m)_0 \qquad \nabla_m J^m = 0 \qquad (10)$$

The theory allows adding some polar vector to $\Gamma_n$. This vector can be interpreted as a vector potential of the electromagnetic field and, hence, can introduce this field into the theory ( see [5]). It is seen from equation (8) that the spin curvature $K_{ik}$ has a potential and vortex part, similar to the electromagnetic field $F_{ik}$. The vortex part describes the whirl-like gravitational structures which can exist in the empty space because $(R_{iklm})_0 \neq 0$ and $(K_{ik})_0 = (1/4)(\sigma^{lm} R_{iklm})_0 \neq 0$.

The linear approximation of the Einstein equation $R_{km} = 0$ for empty space, in the assumption of a small deviation of the metric from a non-perturbed one, describes gravitational waves [2]. Gravitational wave superposition can represent complicated space-time structures. Interactions of gravitational waves can form gravito-solitons [6], i.e. nonlinear structures analogues to solitons in an ordinary liquid. Equation (8) has no linear approximations with respect to tetrads and does not describe gravitational waves. However, it may describe other gravitational structure such as whirl-like structures.

It is known that whirl-like gravitational structures exist in space around heavy rotating bodies (Lense-Thirring effect), and may also be a property of empty space geometry.

All the above equations do not contain the Plank's constant: thus they are classical equations. Matrix equations are possible, though not common, in classical theories. The matrix $\sigma^{ik}$, by itself, does not describe the spin properties of a test particle. This matrix is determined by space geometry, which depends on the distributions and motions of all bodies around.

It is also important to note, that two arbitrary space-time intervals $ds_1$ and $ds_2$ do not commute:

$$ds = \gamma_n dx^n \qquad [ds_1, ds_2] = \sigma_{ik} dx^i \times dx^k \qquad ds^2 = g_{ik} dx^i dx^k \qquad (11)$$

$dx^i \times dx^k = -dx^k \times dx^i$ is the infinitesimal surface element determined by the vectors of displacement. The quantities $ds_r = \sqrt{ds^2}$ can be measured with arbitrary precision with rods and clocks. Final results of the theory, which can be compared with observations, have to be taken in the form which is independent of the $\gamma^n(x)$ representation. For this purpose, square roots from quadratic forms or traces of resulting expressions can be taken as final results. In some sense it is similar to the situation of quantum theory, where the wave function is dependent on the $\gamma^\alpha$ representation, but only the square of these functions describes observations. The relations (10) should be important in the future theory of quantum gravitation.

The structure of the gravitational field in Universe is determined by the empty space geometrical relief and by the ordinary matter distribution. The origin of this geometry could be described by the future quantum theory of the Universe development, beginning from the time of "big bang".

## 3. Observations and the theory

Observations of the "dark matter" effect demonstrate only gravitational interaction of the "dark" and ordinary matters. The certain candidates for dark matter particles are not found [1]. This situation and the absence of the common accepted theory of possible empty space structures and their evolution, does not allow presenting a certain explanation of existing observations. But observations, which are difficult to explain in assumption about the dark matter particles, agree with assumption on the almost stationary empty space gravitational structures.

1. One unexplained fact is that Weakly Interactive Massive Particles (WIMP) models of dark matter predict that dark matter ought to clump together gravitationally at all length scales, from galaxies down to much smaller sub-galactic structures. However, this is not what is observed - no dark-matter structures smaller than several hundreds light-years across have been found by astronomers. This is also true for observations that show numerous dark matter sub-halos in the Milky Way [7].

The absence of the dark matter directly in centers of galaxies can be explained, taking in mind that the gravitational force acting on a test particle is determined by the gradient of the metric [2]. The force acting on a non-relativistic ($v \ll c$) test particle with the mass $m$ in a gravitational field with the metric $g_{ik}$ is $\mathbf{f} = -mc^2 \nabla (\ln \sqrt{g_{00}})$, thus if a local gravitational relief is sufficiently smooth, with a small gradient of the metric, it can be interpreted as a local absence of gravitation.

Dark matter particles, as any other particles, should form dense clouds by gravitational attraction, but such clouds were never observed. If "clouds of dark matter" are in fact the empty space structures, with characteristic scales much larger than several hundreds light years, then the metric gradient is small, and it can explain the observed absence of "dark matter" in the central regions of galaxies. It explains also the absence of the "dark matter" in solar system with size of about few light hours. On such scales only luminous matter distribution will determine the influence of gravitation on massive bodies.

2. The existing theory of star formation, including black holes, does not take into account the possible accumulation of hypothetical dark matter particles. This accumulation could be important because there has to be five times more dark matter than ordinary matter.

This problem does not exist if the dark matter effect is a result of the local non flat geometry of the empty space.

3. An analysis of the normal matter at the centers of many galaxies of all shapes and sizes shows that there is always five times more dark matter than normal matter in regions where the dark matter density drops to one-quarter of its central value. This finding goes against expectations because the ratio of dark to normal matter should depend on the galaxy's history - for example, whether it has merged with another galaxy or remained isolated during its entire life time. Mergers should skew the ratio of dark to normal matter on an individual basis, but this is not the case. Central surface matter density of dark matter halo is constant across galaxies with different evolutionary histories [8].

Such observations can be explained if the empty space gravitational structures are much more stable as flexible clouds of ordinary matter and the distribution of the matter tends to adapt itself to the space relief.

4. Other interesting findings that cannot easily be explained by the WIMP model are related to observations of the super cluster MACSJ0025 that was created in the process of two galactic clusters collision. It was found that the velocity of ordinary matter decreases, but the dark matter velocity remains unchanged. The dark matter has a form of two large clouds. In MACSJ0025 cluster the ordinary and dark matter clouds appear separated in space. The similar separation of dark and ordinary matter was observed in the "Bullet cluster" as well. It appears as if dark matter forms a gravitational scaffold into which gas can be accumulated, and stars can be built [9].

These observations can be considered as an additional evidence of the different origin and shapes of the almost stationary empty space structures and flexible distributions and behavior of ordinary matter in the considered regions. The time scale necessary for normal matter to concentrate in gravitational ravines of a background space relief may be larger than the time scale of cluster collisions. For regions in space where the time scale of matter concentration in gravitational valleys is smaller than time scale of cluster collisions, the ordinary matter tends to concentrate predominantly in gravitational depression of the empty space.

5. The density distribution of the dark matter around spiral galaxies was observed. Their rotation curves, co-added according to the galaxy luminosity, conform to an universal spiral galaxy profile, which can be represented as the sum of exponential thin disk term plus a spherical halo term with a density core. The fine structure of the dark matter halos was obtained from the kinematics of a number of suitable low-luminosity disk galaxies. The halo circular velocity increases linearly with radius out to the edge of the stellar disk, implying a constant dark halo density over the entire disk region. The properties of halos around normal spirals provide substantial evidence of a discrepancy between the mass distributions predicted in the "Cold Dark Matter" scenario and those actually detected around galaxies.

Observations show that the mean dark matter surface density within one scale-length of the dark halo is constant across a wide range of galaxies. The luminous matter surface density is also reported to be constant within one scale-length of the dark halo, such that within one halo scale-length the luminous-to-dark matter ratio is constant, while the total luminous-to-dark matter ratio is not constant among different galaxies [10].

It can be considered as additional evidence that the luminous matter distribution tends to correlate with the local empty space profiles. Shapes of the space structures are similar but not identical each other.

6. Data from the Hubble telescope shows a large population of small galaxies in Perseus galaxy cluster that have remained intact while larger galaxies around them are being ripped apart by the gravitational pull of neighboring galaxies. The dwarf galaxies have even higher amounts of dark matter than spiral galaxies like our Milky Way. It shows that "dark matter" provides scaffolding for the Universe, forming a framework for the formation of galaxies through gravitational attraction [11].

These observations can be explained as evidence of the different depths of the empty space depressions. The small depression could be deeper than the larger ones.

## 4. Conclusion

It follows from equations (1),(3),(4),(8) that the empty space structures are much more complicated than it follows from the often used condition $R_{il} = 0$.

The empty space geometry is determined by non zero values of the Riemann tensor. In particular, this tensor can be dependent on the combination of second rank tensors $\Lambda_{ik}$ which are, in a mathematical sense, identical to energy momentum tensors $T_{ik}$ of the ensemble of particles, but determine the properties of the empty space (physical vacuum, quintessence etc). That means that the gravitational field described by $\Lambda_{ik}$ tensors is the same which can be created in the ensemble of particles and is determined by $T_{ik}$. But then the analogy stops. Ordinary particles interact one with another not only by gravitational force. In opposite, the macroscopic vacuum structures demonstrate the gravitational fields only.

Hence, all observed phenomena which were explained as the demonstration of the dark particles gravitational fields and were described by the energy momentum tensor of the particles $T_{ik}$, (such as galaxy rotation curves, gravitational lensing etc.) can be considered as a consequence of a special empty space structure described by the Riemann tensor .

The Riemann tensor or both, the Ricci tensor and the spin-curvature tensor, are necessary for the complete description of space geometry.

The space geometry and the matter distribution and motion are mutually dependent. All evidences show that the "dark matter" and ordinary matter distributions are correlated, but this correlation is not perfect. This correlation is not similar to the correlation of two types of matter, if they were to consist of any sorts of particles because motions of ordinary particles are determined not only by the space metric but also by non-gravitational forces. The matter, which forms stars and galaxies, concentrates in gravitational ravines and depressions of the empty space. The gravitation of ordinary matter provides perturbations of space geometry.

The large scale (several of Mpc) empty space geometry can describe the dark energy effect. The existence of small scale (several tenth of Kpc) empty space structures can explain the "dark matter" observations.

The $T_{km}$ values for ordinary matter at different points in space can be determined only by observations. The same is true for $(R_{iklm})_0$ Equations (1), (3), (8) determine $h_\alpha^k(x)$, and hence $g_{lm}$ and $\sigma_{ik}$. Gravitational forces acting on a particle in space can be described in terms of the metric $g_{il}$, using the well known formulas of general relativity [2].

Our explanation of the "dark matter" is based on assumptions that the empty space geometry has many local in-homogeneities with different scales including the scales on the order of the galaxy size. This assumption: a) does not contradict observations, b) is based on the Riemann geometry,  c) is much less arbitrary, compared to existing assumptions about "dark matter" particles with extraordinary properties.

Our results do not contradict the general ideas of the inflation theory. We do
not discuss here the origin of the empty space geometrical structure which should be possible by the future theory of  quantum gravitation.